\documentclass[twocolumn,showpacs,preprintnumbers,amsmath,amssymb]{revtex4}
\usepackage{graphicx,color}
\usepackage{dcolumn}
\usepackage{bm}

\begin{document}

\preprint{}

\title{Shape coexistence near the neutron number N$=$ 20:\\ 
       First identification of the E0 decay from the deformed 
       $0^+_2$ state in $^{30}$Mg}

\author{W.~Schwerdtfeger$^1$, P.G.~Thirolf$^1$, K.~Wimmer$^1$, D.~Habs$^1$,
	H.~Mach$^2$, T.R.~Rodriguez$^4$,
	V.~Bildstein$^3$,
        J.L.~Egido$^4$,
	L.M.~Fraile$^5$,
	R.~Gernh\"auser$^3$,
        R.~Hertenberger$^1$,
	K.~Heyde$^6$, P.~Hoff$^7$,
	H.~H\"ubel$^8$,
	U.~K\"oster$^9$,
	T.~Kr\"oll$^3$,
	R.~Kr\"ucken$^3$,
	R.~Lutter$^1$,
	T.~Morgan$^1$ and
	P.~Ring$^3$}
\affiliation{$^1$Ludwig-Maximilians-Universit\"at M\"unchen, D-85748 Garching, Germany \\
$^2$Department of Nuclear and Particle Physics, Uppsala University, SE-75121 Uppsala, Sweden \\
$^3$Physik Department E12, Technische Universit\"at M\"unchen, D-85748 Garching, Germany \\
$^4$Universidad Autonoma de Madrid, E-28049 Madrid, Spain \\
$^5$Universidad Complutense, E-28040 Madrid, Spain \\
$^6$Department of Subatomic and Radiation Physics, Universiteit Gent, B-9000, Gent, Belgium \\
<$^7$Department of Chemistry, University of Oslo, N-0315 Oslo, Norway \\
$^8$Helmholtz-Institut f\"ur Strahlen- und Kernphysik, Universit\"at Bonn, D-53115 Bonn, 
    Germany \\
$^9$Institut Laue-Langevin, F-38000 Grenoble, France}

\date{\today}

\begin{abstract}

The 1789~keV level in $^{30}$Mg was identified as the first
excited $0^+$ state by measuring its E0 transition to the ground state.
The measured small value of $\rho^2($E0$,0^+_2 \rightarrow 0^+_1) = 5.7(14) 
\cdot 10^{-3}$ implies a very small mixing of competing configurations
with largely different intrinsic quadrupole deformation near N$=$20.
Axially symmetric Beyond-Mean-Field configuration mixing calculations
identify the ground state of $^{30}$Mg to be based on neutron configurations 
below the $N=20$ shell closure, while the excited 0$^+$ state mainly 
consists of a two neutrons excitated into the $\nu~1 f_{7/2}$ orbital.
Using a two-level model, a mixing amplitude of 0.08(4) can be derived.

\end{abstract}

\pacs{23.40.-s,23.20.Nx,23.20.Js,27.30.+t}

\maketitle

One of the most studied phenomena in the region of neutron-rich atomic nuclei 
around the $N=20$ shell closure is the occurrence of strongly 
deformed ground states in Ne, Na and Mg isotopes. This so-called 
'island of inversion'~\cite{war90} (discovered
already more than 30 years ago~\cite{thibault75}) finds its origin in the 
promotion of a pair of neutrons across the $N=20$ shell gap, thus leading 
to the intrusion of deformed low-lying (2p2h) configurations below the 
spherical (0p0h) states compared to nuclei closer to $\beta$ stability. 
Despite considerable 
efforts the precise 
localization of the transition from normal to intruder-dominated 
configurations is not yet finally settled and even the origin of the large 
collectivity of the $0^+_{gs}\rightarrow 2_1^+$ transition in $^{32}$Mg is 
still under debate~\cite{yama04}.
A coexistence of spherical and deformed $0^+$ states is predicted to
exist within a small region around N$=$20 in the neutron-rich Mg 
nuclei~\cite{heyde,wood92}, the 'island of inversion'.
So far studies on spectroscopic properties have focused on B(E2) values 
between the 0$^+$ ground state and the first excited 2$^+$ 
state~\cite{moto95,iwasaki01,prity99,chiste01,nied05,nied05a,Gad07}, however 
no excited 0$^+$ state has been experimentally observed in these nuclei.
While $^{32}$Mg, which is conventionally considered a closed-shell nucleus,
 exhibits a strongly deformed 
ground state as indicated by the large value of 
B(E2;$0^+_{gs}\rightarrow 2^+_1$)=454(78)e$^2$fm$^4$~\cite{moto95}, the 
ground state of $^{30}$Mg is expected to be much less deformed, whereas a 
(deformed) excited $0^+_2$ state is predicted at an energy between 
1.7 and 2~MeV~\cite{RodGuz02,Caur01,Nov02,Ots04}
that has so far escaped observation.

It is the purpose of this Letter to report the unambiguous identification
of the coexisting $0_2^+$ state in $^{30}$Mg using conversion electron
spectroscopy and to discuss the configuration mixing between normal and 
intruder configurations at the border of the 'island of inversion'.
 
\begin{figure}
	\includegraphics[width=0.48\textwidth]{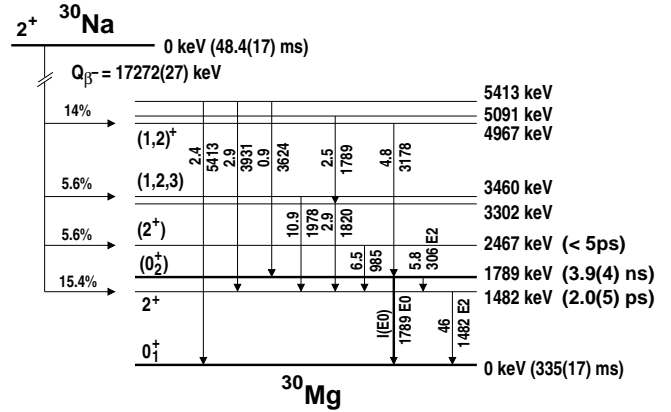}
	\caption{Low-energy part of the level scheme of 
                 $^{30}$Mg~\cite{Mach05}.}
	\label{fig:niveau}
\end{figure}

\begin{figure}
	\includegraphics[width=0.35\textwidth]{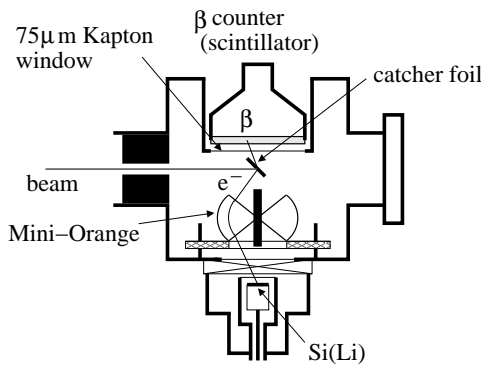}
	\caption{Sketch of the experimental setup (top view). The Germanium 
                 detector that was mounted vertically on top of the target chamber 
                 is not shown.}
	\label{fig:setup}
\end{figure}

Resulting from fast timing $\gamma$-spectroscopy studies~\cite{Mach05}, 
the 1789~keV level in $^{30}$Mg emerged as a strong candidate for the 
deformed first excited 0$^+_2$ state due to its long half-life of 3.9(4)~ns 
and the absence of a ground state $\gamma$ transition. Fig.~\ref{fig:niveau} 
displays our present knowledge on the low-energy part of the level scheme of 
$^{30}$Mg~\cite{Mach05}, triggering our search for the deformed
0$^+_2$ state in $^{30}$Mg via conversion electron spectroscopy following 
$\beta$ decay of $^{30}$Na at the ISOLDE facility at CERN.
The radioactive $^{30}$Na atoms [$t_\frac{1}{2} = 48.4(17)$~ms] were 
produced by
sending 1.4~GeV protons provided by the CERN PS Booster with an intensity up 
to $3.2 \cdot 10^{13}$ p/pulse onto a UC$_x$/graphite target (heated to $\sim$ 2050$^o$C).
On average every second pulse (repetition rate of 1.2~s) 
was used. The reaction products diffusing out of the target were 
surface-ionized and the extracted $1^+$ ions were mass separated by the 
ISOLDE High Resolution Separator at a kinetic energy of 40~keV. 
This $A = 30$ beam was sent to the experimental setup which is illustrated 
in Fig.~\ref{fig:setup}.
The beam was stopped in the center of the target chamber in a 0.1~mm thick
Al foil to examine the $\beta$ decay of $^{30}$Na to excited states of
$^{30}$Mg. In order to detect the E0 decay electrons with high
resolution (3.0~keV FWHM) a liquid nitrogen cooled Si(Li) detector with an active
surface of 500~mm$^2$ and a thickness of 5~mm was used.
The detector was
operated in conjunction with a Mini-Orange~\cite{vklinken} consisting of 8 
wedge-shaped (Nd$_2$Fe$_{14}$B) permanent magnets creating a toroidal 
magnetic field (B$\sim$160~mT) arranged around a central Pb absorber 
(diameter: 16~mm, length: 50~mm) in order to block $\gamma$ rays from the 
catcher foil. Towards the catcher foil the absorber was covered by a Cu cap 
in order to suppress X-ray production. The transmission efficiency of the
spectrometer was 3.2~\% around 1.7~MeV, optimized for the expected E0
transition in $^{30}$Mg. A 0.2~mm thick plastic scintillator
(BC-404, diameter 50~mm) read out by a 2'' photomultiplier tube was mounted 
at a distance of
13~mm to the target, resulting in a solid angle coverage of 
$\Omega/4\pi = 21~$\%. This detector served as trigger on $\beta$-decay 
electrons and was operated in coincidence with the Mini-Orange spectrometer. 
In order to identify the beam composition and for normalisation purposes 
$\gamma$ rays following the $\beta$ decay were
detected using a Ge detector mounted on top of the target chamber.

In order to achieve optimum sensitivity for the expected weak E0 transition, 
the setup was optimized for the reduction of the potentially dominant 
coincident background from Compton electrons in view of the large 
Q value of the $^{30}$Na $\beta$ decay (17272(27)~keV). Consequently the Al 
target 
chamber was coated at the inside by 15~mm thick Teflon plates in order to 
absorb Compton-scattered electrons.
The germanium detector served as monitor of the $A=30$ beam composition, 
which turned out to consist almost entirely of $^{30}$Na at a total intensity 
of 4100 decays/second. 
Since the halflife of $^{30}$Na decay is much shorter compared to 
$^{30}$Mg or $^{30}$Al originating from the $\beta$ 
decay chain, during the analysis events with short 
lifetimes ($\leq$ 400~ms) were selected from a decay time measurement relative 
to the initial proton pulse, thus enhancing spectral contributions 
from $^{30}$Na.

\begin{figure}
	\includegraphics[width=0.45\textwidth]{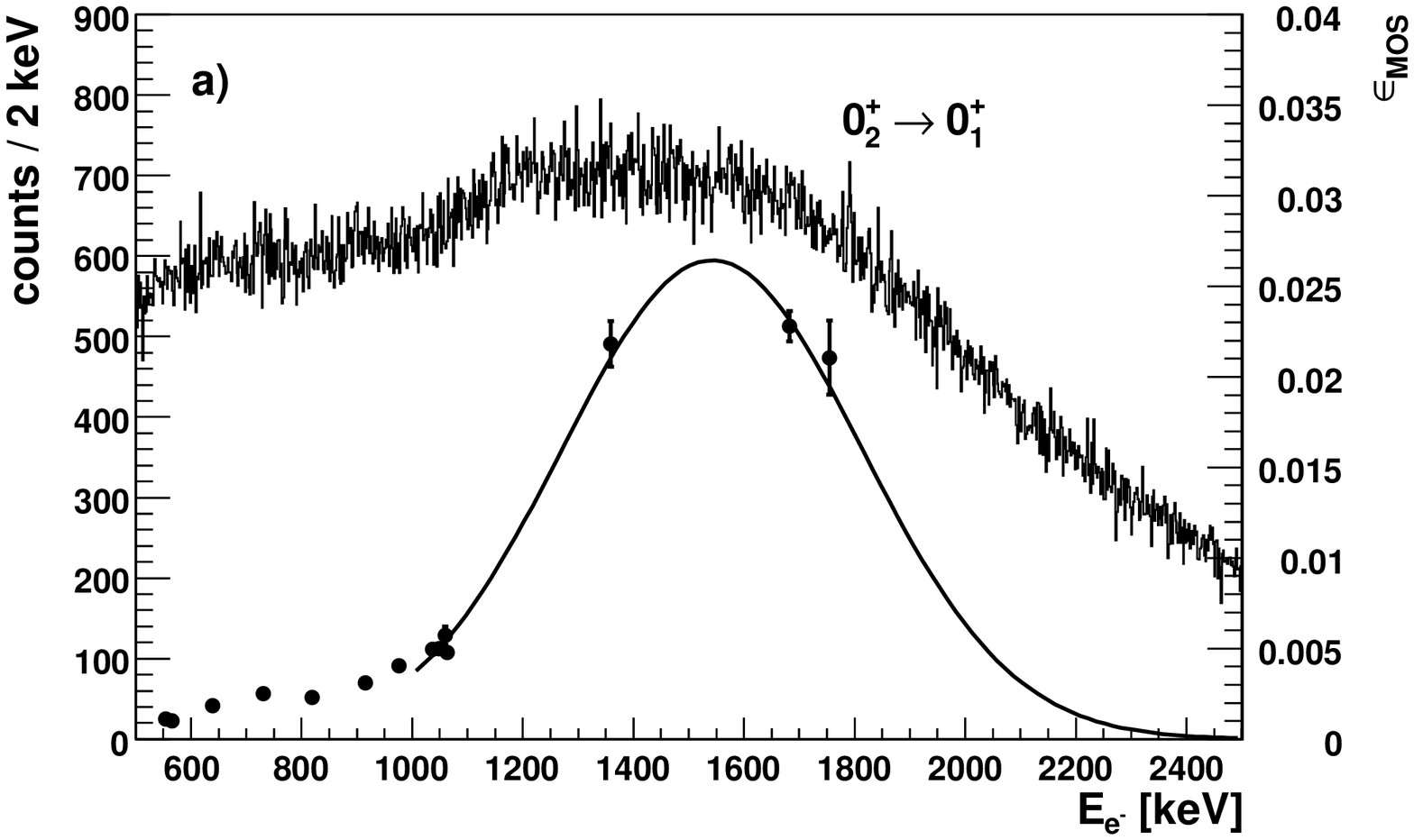}
	\includegraphics[width=0.45\textwidth]{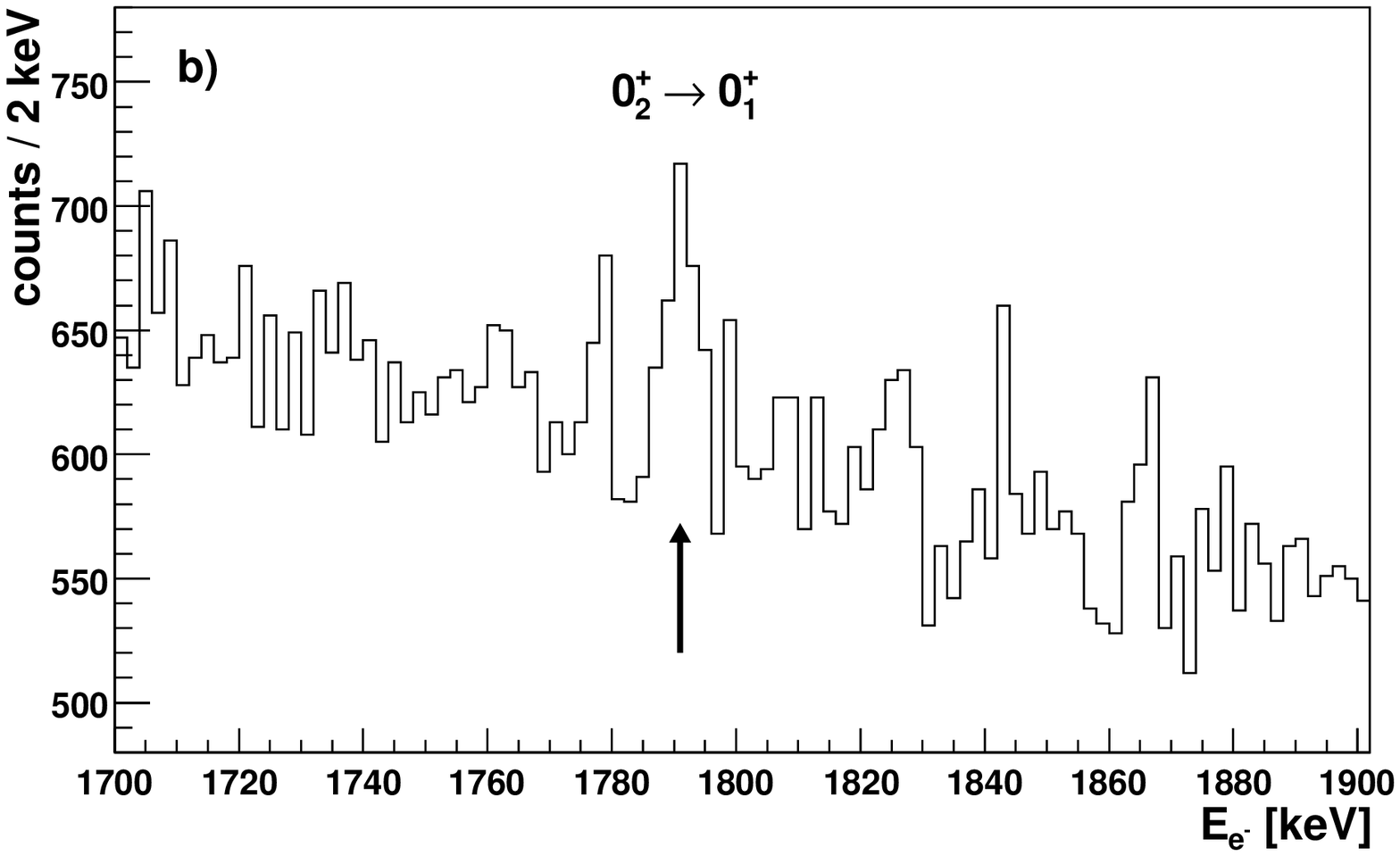}
	\caption{a) Background-subtracted electron spectrum measured in 
                 coincidence with signals in the plastic detector, gated on
                 time differences $\leq$ 500~ms between the proton pulse and
                 the registered $\beta$ decay signal. Also shown is the 
                 transmission efficiency curve of the Mini Orange 
                 spectrometer. The 
                 rapidly increasing $\beta$ decay energy spectrum 
                 (Q$_{\beta}=$17.3~MeV) gives rise to the rather high background 
                 remaining below the transmission maximum.
                 b) Expanded view to the electron spectrum in the region of 
                 the E0 transition in $^{30}$Mg.}
	\label{fig:E0}
\end{figure}

The resulting electron spectrum detected with the Si(Li)
detector in coincidence with $\beta$-decay electrons is shown in 
Fig.~\ref{fig:E0}. The upper panel a) displays the spectrum over a wide 
energy range, together with the transmission efficiency curve of the 
Mini-Orange 
spectrometer (solid line). The rather high yield below the transmission 
maximum results from the rapidly increasing $\beta$ decay energy 
spectrum. In Fig.~\ref{fig:E0}b) the sought $0^+_2 \rightarrow 0^+_1$ E0
transition at 1788~keV is visible. 304(60) counts were detected in the
peak during 143~hrs of beamtime (MO transmission efficiency: 2.0(2)$\%$).
With an energy resolution of 3.0~keV for the Si(Li) detector spanning K and L 
conversion in $^{30}$Mg (E$_K=$1.3~keV), the monopole strength 
$\rho^2($E0$)$ can be determined by the ratio of E0 (K+L) conversion 
intensity to the E2 $\gamma$ intensity and the $\gamma$ lifetime 
$\tau_{\gamma}$ of the 0$_2^+$ state as~\cite{kibedi05}

\begin{equation}
   \rho^2({\rm E0}) = \frac{I_{\rm K+L}({\rm E0})}{I_\gamma({\rm E2})} \cdot
	\frac{1}{\Omega_{\rm K+L}({\rm E0})} \cdot \frac{1}{\tau_{\gamma}}.
\end{equation}

The $\gamma$ intensity of the E2 transition at 306~keV measured within
the $\beta$ - $\gamma$ coincidence condition using the Ge detector is
$I_\gamma($E2$) = 6.591\cdot 10^4$ ($\epsilon_\gamma = 0.0019$), the half-life
$T_{1/2}$ of the $0^+_2$ state was measured to be $3.9(4)$~ns~\cite{Mach05}
(corresponding in our case to $\tau_{\gamma}\approx\tau =$5.6(6)~ns) 
and 
the electronic $\Omega$ factor is $\Omega_{K+L} = 1.39 \cdot 10^7$/s 
\cite{kant95}. This finally allows to derive the electric monopole matrix 
element $\rho^2($E0,$^{30}$Mg$) = 5.7(14) \cdot 10^{-3}$, which corresponds 
to an intensity of the E0 transition of $I=2.2(5)\cdot 10^{-5}$ and a 
partial E0 lifetime of $\tau(E0) = 1.26(31)~\mu$s.

In the 'island of inversion' the deformed configuration based on two neutrons 
being excited from the $\nu~1 d_{3/2}$ to the intruder orbital $\nu~1 f_{7/2}$
keeps pace with the normal spherical one as illustrated by the 
case of $^{32}$Mg, where the intruder state even becomes the ground state. 
In such a situation of competing configurations and in the absence of mixing 
one expects either a deformed $0^+_1$ and a nearly spherical  $0^+_2$ state 
or the other way around. Since the E0 operator is a {\it single particle} one, 
one expects in both cases small values of the monopole matrix element 
$M(E0)\equiv \langle 0_g^+ \vert T(E0) \vert 0_{\rm exc}^+ \rangle $
with T(E0) being the E0 operator as given by Wood et al.~\cite{wood99}
and $\rho^2(E0) \equiv \vert {M(E0)/eR^2} \vert ^2$ with the nuclear 
radius $R=1.2~A^{{1}/{3}}$ fm.
Inducing configuration mixing, somewhat larger values of 
$\rho^2({\rm E0})$ can be expected. In $^{30}$Mg the small experimental 
value of $\rho^2($E0$) = 5.7(14) \cdot 10^{-3}$ points towards the presence 
of small mixing. However, an important question that remains to be 
answered concerns the
nature of the two $0^+$ states and the amount of mixing of 
the $\nu~1 d_{3/2}$ and $\nu~1 f_{7/2}$ configurations. Concerning the 
$0^+_1$ state there are  strong experimental~\cite{prity99,nied05} and 
theoretical \cite{Caur01,RodGuz02,Ots04} indications that in $^{30}$Mg the 
inversion has not taken place.

In order to understand the experimental findings
calculations going beyond the mean-field by incorporating configuration
mixing~\cite{rod07} have been 
performed using the finite range density dependent Gogny force with the 
D1S parameterization~\cite{ber84}.
The results of these calculations for $^{30}$Mg are listed and compared 
to experimental values in Table~\ref{tab:egido}. The excitation energy of 
the 0$_2^+$ state and the B(E2;$0_1^+\rightarrow 2_1^+$) value agree 
reasonably well with the experimental values, while the value of 
$\rho^2(E0)$ overestimates the experimental result.

\begin{table}
 \begin{center}
   \begin{tabular}{c|c|c|c|c}
     & E$_x$($2_1^+$) & E$_x$($0_2^+$)  & B(E2, $0_1^+ \rightarrow 2_1^+$) 
                                       & $\rho^2$(E0) $\cdot 10^3$ \\
     &  (MeV)& (MeV) &  ($e^2 fm^4$)     &    \\
   \hline
   Theory      &  2.03   &  2.11  &  334.6                 &  46      \\
   Experiment  &  1.482  &  1.789 &  241(31) \cite{nied05} &   5.7(14) 
   \end{tabular}
   \caption{Results from Beyond Mean Field calculations with Gogny force 
            for $^{30}$Mg compared to experimental values.}
   \label{tab:egido}
 \end{center}
\end{table}

\begin{figure}
   \includegraphics[width=0.5\textwidth]{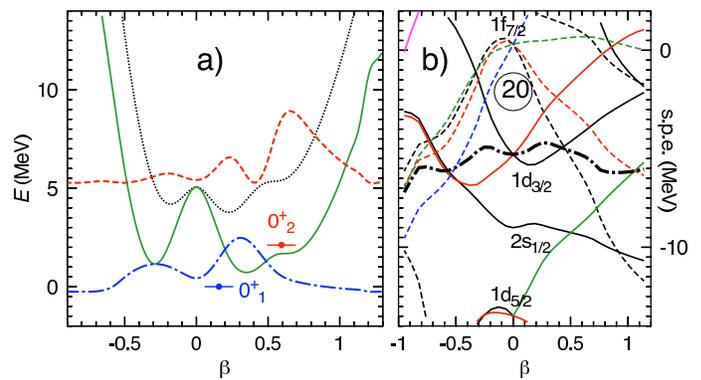}		
   \caption{Theoretical results for $^{30}$Mg: a) The dotted line corresponds 
            to results of particle number projected (PNP) calculations, the 
            green full line to the $J=0$ energy 
            surface and the blue (red) dashed-dotted (dashed) line 
            represents the collective 
            wave function of the $0^+_1$ ($0^+_2$) state. b) Neutron 
            single-particle energies as a function of deformation. 
            The thick dash-dotted line represents the Fermi level (color).}
   \label{fig:spe-wf}
\end{figure}

In Fig.~\ref{fig:spe-wf}a, the particle number projected (PNP) 
potential energy curve displays 
both a mild prolate and an oblate minimum at small deformation and a 
shoulder at 
larger $\beta$ values. Inspecting the neutron single-particle energies
in the right hand panel, we see that the two minima at moderate deformation 
correspond to the two minima of the single-particle energies of the 
$\nu~1 d_{3/2}$ orbitals just below the Fermi level (notice that this is not 
the case in $^{32}$Mg, see~\cite{RodGuz02}), while the shoulder appears at 
deformations at which two neutrons already occupy the $\nu~1 f_{7/2}$ orbital. 
The angular momentum projection provides an additional energy lowering with 
respect to the PNP energy (the full line in panel (a)), and finally 
configuration mixing leads to the $0^+_1$ and $0^+_2$ states positioned in 
the E-$\beta$ plane (Fig.~\ref{fig:spe-wf}a) according to their energy and average 
deformation. The 
composition of the collective wave functions of these two states, 
i.e. the weights of the corresponding $\beta$ values being admixed, 
indicates the character of the state. We notice that the $0^+_1$ state 
(blue dotted-dashed line) is a mixture 
of prolate and oblate $\nu~1 d_{3/2}$ configurations which average to a small 
intrinsic deformation of $\beta=0.16$. The $0^+_2$ state (blue dashed line), 
on the other hand, 
is a well deformed state with $\beta=0.59$ consisting to a large part of 
a $\nu~1 f_{7/2}$ configuration with very small admixtures of the 
$\nu~2 d_{3/2}$ 
configuration. These results are consistent with the experimental 
finding of a very small matrix element connecting both 0$^+$ levels.

It is known~\cite{bouch03,petro00} that the calculation of $\rho^2(E0)$ is 
very sensitive to small variations of the interaction matrix elements and 
in particular to small admixtures of different shapes. 
Ideally, one needs to consider triaxial shapes in the calculations.
Due to the complexity of a fully triaxial angular momentum projected
calculation, at present, we had to impose restrictions to axial symmetric
states only. Neglecting triaxial effects might be a worse approximation 
in the calculation of $\rho^2(E0)$ than in the case of other observables like
energies or B(E2) strengths, because in the latter cases the intrinsic state 
of the initial and final wave functions is the same, while in the case of
$\rho^2(E0)$ these states, in general, have a rather different dependence 
on $\gamma$. So there is room for possible improvement over the present
theoretical $\rho^2(E0)$ value.

In order to quantify the mixing amplitude between the deformed and spherical
configurations, we have also analyzed our experimental results making
use of a phenomenological two-level mixing model. Here the monopole 
matrix element 
 strongly depends on the mixing amplitude $a$ between the two 
intrinsic (spherical and deformed) $0^+$ states, i.e.

\begin{eqnarray}
   |0^+_g \rangle & = & \sqrt{1-a^2} |0^+_{\rm sph} \rangle + a |0^+_{\rm def} 
    \rangle~, \\
   |0^+_{\rm exc} \rangle & = & - a |0^+_{\rm sph} \rangle + \sqrt{1-a^2} 
   |0^+_{\rm def}  \rangle~. \nonumber
   \label{eq:expSyst}
\end{eqnarray}

Within the quadrupole deformed rigid rotor model the strength of the E0 
transition can be described as~\cite{wood99}

\begin{equation}
   \rho^2(E0)=(\frac{3}{4\pi} Z)^2 \cdot a^2\cdot (1-a^2)\cdot
   (\beta_1^2-\beta_2^2)^2~,
  \label{eq:amplitude}
\end{equation}

with $\beta_1, \beta_2$ the equilibrium quadrupole deformations
corresponding to the $|0^+_{\rm sph}\rangle$ and $|0^+_{\rm def}\rangle$ 
intrinsic configurations, respectively.
Using the deformation values of the two $0^+$ states as calculated above 
($\beta_1 (0_1^+) =$0.16, $\beta_2 (0_2^+) =$0.59) together with the 
experimental value of $\rho^2(E0)$, a value of $a^2 =$ 0.0067(16) 
can be extracted, resulting in a value of $a= 0.08(4)$ for the mixing 
amplitude between the two intrinsic 0$^+$ states.
It should be noted that an identical value of $\beta_2 (0_2^+) =$0.59
can be inferred from the phenomenological Grodzins 
systematics~\cite{raman01}, which empirically correlates the 
B(E2;$0_1^+ \rightarrow 2_1^+$) value and the excitation energy of the 
first excited 2$^+$ state (here based on the assignment of the 2467~keV 
level as being the rotational $2_2^+$ state). For $^{30}$Mg and $^{32}$Mg 
the B(E2) values predicted by the Grodzins systematics (256(45)~e$^2$fm$^4$ 
and 410(73)~e$^2$fm$^4$) agree remarkably well with the experimental values 
(241(31)~e$^2$fm$^4$~\cite{nied05} and 454(78)~e$^2$fm$^4$~\cite{moto95}). 
This also points to rather pure $0^+$ and $2^+$ states in the two potential 
minima, because otherwise deviations from the Grodzins systematics could 
be expected.

To conclude, conversion electron measurements at ISOLDE have identified 
the 1789~keV level 
as the $0^+_2$ state in $^{30}$Mg. The conversion electrons were
measured in coincidence with $\beta$ decay electrons using a Mini-Orange
spectrometer. The monopole
strength extracted from measuring the $0^+_2 \rightarrow 0^+_1$ E0 transition
is $\rho^2($E0,$^{30}$Mg$) = 5.7(14)\cdot 10^{-3}$, which corresponds 
to an intensity of the E0 transition of $I=2.2(5)\cdot 10^{-5}$ and a 
partial E0 lifetime of $\tau(E0) = 1.26(31)~\mu$s.
The small value of the monopole strength indicates a weak mixing and allows 
for the first time to 
deduce the mixing amplitude between shape-coexisting 0$^+$ states near the 
'island of inversion' as $a = 0.08(4)$.
State of the art beyond-mean-field calculations identify the ground 
state as based on a mixture of prolate 
and oblate $\nu~1 d_{3/2}$ orbitals and the excited $0^+_2$ state on a rather 
pure $\nu~1 f_{7/2}$ level, thereby confirming a sharp borderline of the 
'island of inversion' and a very weak mixing between the competing 
configurations.

\begin{acknowledgments}
  This work was supported by BMBF (contracts 06ML234, 06MT238 and
  06BN109), by the European Commission within FP6 through I3-EURONS
  (contract no. RIDS-CT-2004-506065), by the Swedish Research Council,
  by MEC (FPA2007-66069) and by 
  the Spanish Consolider-Ingenio 2010 Program CPAN (CSD2007-00042). 
\end{acknowledgments}

\end{document}